\documentclass[usenatbib,useAMS]{mn2e}

\usepackage{times}
\usepackage{rotating}
\usepackage{graphicx}
\usepackage{epstopdf}
\usepackage{amsmath}
\usepackage{amssymb}
\usepackage{bbold}
\usepackage{multirow}
 
\usepackage{xcolor}

\usepackage{float}

\newcommand{\mbi}[1]{\mbox{\boldmath$#1$}}

\newcommand{\mat}[1]{\mbox{\rm\bf #1}}
\newcommand{\lsim}{\mbox{${\,\hbox{\hbox{$ < $}\kern -0.8em \lower 1.0ex\hbox{$\sim$}}\,}$}}
\newcommand{\gsim}{\mbox{${\,\hbox{\hbox{$ > $}\kern -0.8em \lower 1.0ex\hbox{$\sim$}}\,}$}}

\def\beqn{\vspace{2mm}
\begin{eqnarray}} 
\def\eeqn{\vspace{2mm} 
\end{eqnarray}}

 \def\la{\langle}

\newcommand{\be}{\begin{equation}}
\newcommand{\ee}{\end{equation}}
\newcommand{\ba}{\begin{eqnarray}}
\newcommand{\ea}{\end{eqnarray}}
\newcommand{\brr}{\begin{array}}
 
\newcommand{\err}{\end{array}}
\newcommand{\bc}{\begin{center}}
\newcommand{\ec}{\end{center}}

\begin{document}
\title[The Initial Conditions]{The Initial Conditions of the Universe from Constrained Simulations}

\author[F.~S.~Kitaura]{Francisco-Shu Kitaura\thanks{E-mail: kitaura@aip.de, Karl-Schwarzschild-fellow}\\
Leibniz-Institut f\"ur Astrophysik Potsdam (AIP), An der Sternwarte 16, D-14482 Potsdam, Germany}

\maketitle

\begin{abstract}

I present a new approach to recover the {\color{black} primordial density fluctuations}  and the cosmic web structure underlying a galaxy distribution.
The method is based on sampling Gaussian fields which are compatible with a galaxy distribution and a structure formation model. This is achieved by splitting the inversion problem into two Gibbs-sampling steps: the first being a Gaussianisation step transforming a distribution of point sources at Lagrangian positions --which are not a priori given-- into a linear alias-free Gaussian field. This step is based on Hamiltonian sampling with a Gaussian-Poisson model. The second step consists on a  {\color{black} likelihood comparison} in which the set of matter tracers at the initial conditions is constrained on the galaxy distribution and the assumed structure formation model. For computational reasons second order Lagrangian Perturbation Theory is used.  {\color{black} However, the presented approach is flexible to adopt any structure formation model. }
A  semi-analytic halo-model based galaxy mock catalog is taken to demonstrate that the recovered initial conditions are closely unbiased with respect to the actual ones from the corresponding $N$-body simulation down to scales of a {\color{black} $\sim$5 Mpc/$h$}. The cross-correlation between them shows a substantial gain of information, being at $k\sim0.3$ $h$/Mpc more than doubled.
In addition the initial conditions are extremely well Gaussian distributed and the power-spectra follow the shape of the linear power-spectrum being very close to the actual one from the simulation down to scales of $k\sim1$ $h$/Mpc.
\end{abstract}

\begin{keywords}
(cosmology:) large-scale structure of Universe -- galaxies: clusters: general --
 catalogues -- galaxies: statistics
\end{keywords}

\section{Introduction}

The {\color{black} primordial density} fluctuations of the Universe  comprise all the information of the cosmological large-scale structure at any later time assuming that the theory of structure formation is known.
 Accurate estimates of the initial cosmic density field would hence also lead to a reconstruction of the formation of cosmic structures. This is one of the main motivations underlying constrained simulations of the local Universe \citep[see e.g.][]{mathis,klypin,lavaux}.
The reconstructed density field is especially valueable as it permits one to 
 study the cosmic web  and perform environmental studies  \citep[see e.g.][]{hahn,forero,jasche_sdss,cuartas,wang_sdss,2011MNRAS.416.2494P}.
Moreover, undoing the major effects of gravity on large scales has been shown to  increase the cosmological information.
For this purpose a large amount of techniques (mostly based on local transformations) has been developed  which Gaussianise the cosmic density field
\citep[see e.~g.~][]{2009ApJ...698L..90N,2011ApJ...731..116N,2011arXiv1104.1399J,weinberg,2011arXiv1103.2858Y,2011ApJ...731..116N,2011ApJ...728...35Z}. I refer to \citet[][]{kit_lin} for more complex nonlocal linearisation schemes. 
 In particular tracing the structures back in time within  the Zeldovich approximation \citep{1970A&A.....5...84Z}   can already significantly improve the baryon acoustic oscillation (BAO) measurements  \cite[see e.g.][]{2007ApJ...664..675E,2012arXiv1202.0090P,2012arXiv1202.0091X,2012arXiv1202.0092M}.

The classical approach in the literature to solve the boundary problem to find  the initial positions
of a set of particles governed by the Eulerian equation of motion and gravity is based on the least action principle \citep[see][]{Peebles89,NB00,BEN02}.  A similar
approach consists on relating the observed positions of matter tracers (e.g.
galaxies) in a geometrical way to a homogeneous distribution by minimizing a
cost function \citep[see e.g.][]{LMCTBS08}. One still needs then to find the corresponding Gaussian field to that point source distribution \citep[see e.g.][]{lavaux}.

Ideally, one would wish to sample the density field $\delta(\{\mbi q\})$ at the initial conditions  $\{\mbi q\}$ (Lagrangian positions) given the data at the present, in our case study, given a set of galaxies with their Eulerian coordinates $\{\mbi x_{\rm G}\}$: 
\be
\label{eq:step0}
\delta(\{\mbi q\})\curvearrowleft P(\delta(\{\mbi q\})|\{\mbi x_{\rm G}\} )  \,,
\ee
where the arrow indicates the sampling process.
 The relationship between both coordinate systems is summarised by the following equation:
 $\mbi x=\mbi q+\mbi \Psi$, where $\mbi\Psi$ is the displacement which a particle suffers to go from its initial position $\mbi q$ to its final position $\mbi x$. One should note that the structure formation model is encoded in the displacement field $\mbi \Psi$.
However, this  direct approach is extremely complex since the PDF is highly non-Gaussian as the galaxies {\it live} in structures which  have undergone  nonlinear structure formation. An attempt to find a statistical formulation of such a PDF implies including  the biased nature of a galaxy distribution beyond the Poisson statistics \citep[][]{kitaura_lapalma} and higher order correlation functions beyond the 2-point statistics \citep[][]{kitaura_skewlog}. {\color{black} Such a statistical description could be relevant to deal with primordial non-Gaussianities which will not be considered in this work.}

I suggest in this letter to radically simplify the problem splitting it into well defined ones which encode our physical understanding of structure formation on large scales  in a forward approach.
In the next section, the method is presented in detail followed by a validation section based on tests with synthetic data. Finally the conclusions and discussion section is presented.

\section{Method}






Let us suppose that we have an unbiased sample of matter tracers at initial (Lagrangian) positions $\{\mbi q\}$. {\color{black} The problem of  the reconstruction of the primordial density fluctuations} would be reduced to find the Gaussian field corresponding to a discrete point-source distribution.
Once we knew the initial Gaussian field, we could apply a model of structure formation and match the structures we find in our simulation with the actual ones given by the {\it observed} galaxy distribution  {\color{black} in a likelihood comparison procedure}. 
In this way an iterative procedure can be constructed which yields a set of  Gaussian fields corresponding to the initial distribution of  matter tracers  constrained on some data. The set of solutions will depend on the particular structure formation model and the matching criteria.

In terms of conditional PDFs the two steps described above can be expressed in the following way:
\ba
\label{eq:shot}
\delta(\{\mbi q\})&\curvearrowleft& P(\delta(\{\mbi q\})|\{\mbi q\} ) \\
\label{eq:structure}
\{\mbi q\}&\curvearrowleft& P(\{\mbi q\}|\{\mbi x_{\rm G}\},\delta(\{\mbi q\},m_{\rm SF}) )  \,,
\ea 
where $m_{\rm SF}$ denotes the assumed structure formation model.

\begin{figure}
\includegraphics[width=8.cm]{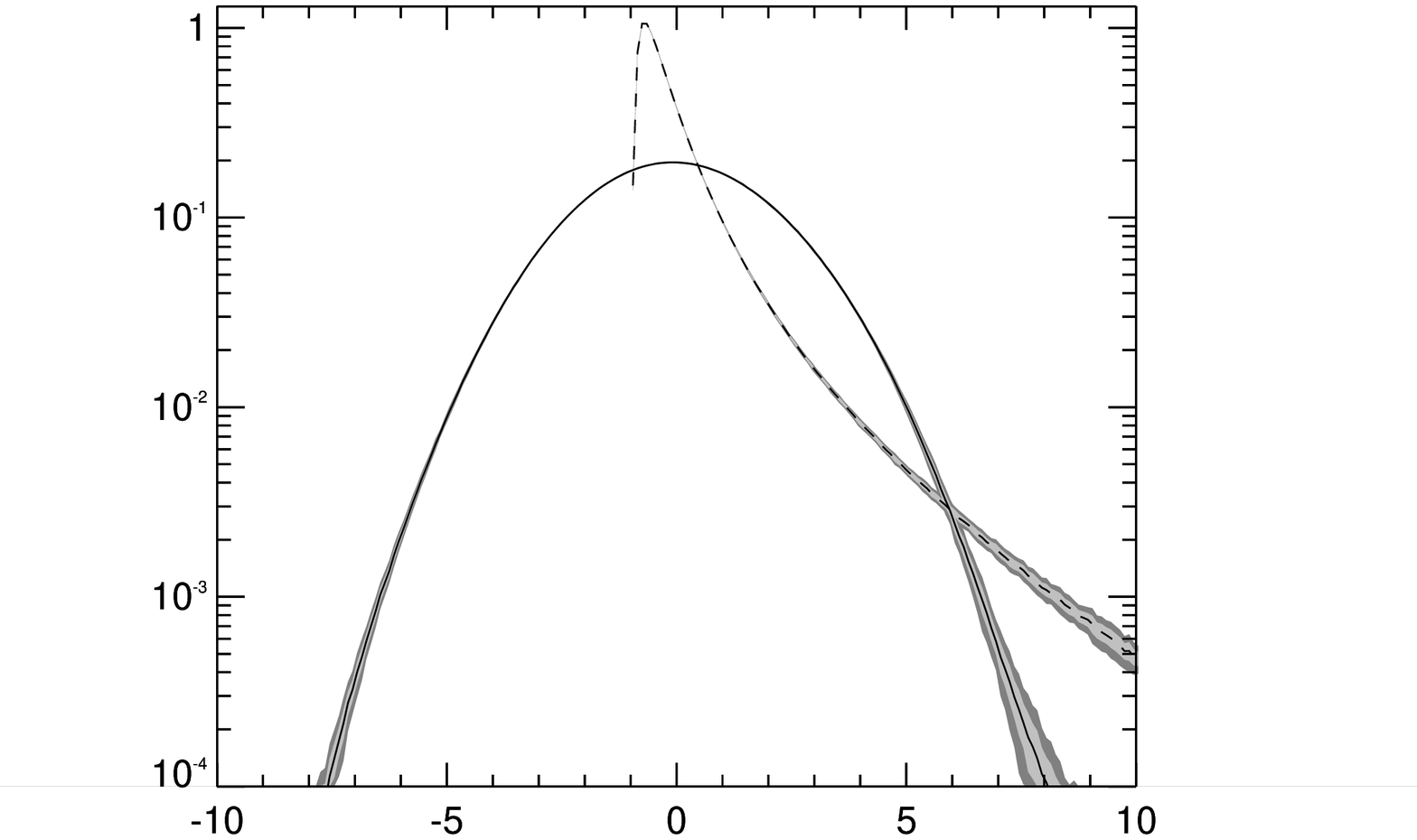}
\put(-235,100){\rotatebox[]{90}{PDF}}
\put(-110,-5){$\delta$}
\put(-203,170){$\langle\mathcal S(\delta(\{\mbi q\}))\rangle\simeq0.05$}
\put(-203,160){$\langle\mathcal K(\delta(\{\mbi q\}))\rangle\simeq-0.03$}
\put(-203,150){$\langle\mathcal M(\delta(\{\mbi q\}))\rangle\simeq-10^{-6}$}
\caption{Matter statistics. Black curve represents the PDF of the overdensity for the mean over 500 samples of the reconstructed initial Gaussian field. Shaded regions indicate 1 and 2 sigma contours. The dashed curve stands for the PDF of the nonlinear 2LPT reconstruction of the same sample. The average skewness, kurtosis and mean of the initial conditions are also indicated by $\langle\mathcal S(\delta(\{\mbi q\}))\rangle$, $\langle\mathcal K(\delta(\{\mbi q\}))\rangle$ and $\langle\mathcal M(\delta(\{\mbi q\}))\rangle$, respectively.} 
\label{fig:stats}
\end{figure}

\begin{enumerate}

\item {\bf Step 1 (Eq.~\ref{eq:shot}): Gaussianisation step: Sampling the Gaussian field given a set of matter tracers at the initial conditions. }  

The first step  can be solved using Bayes theorem. The posterior distribution $P(\delta(\{\mbi q\})|\{\mbi q\} )$ is given by the product of a Gaussian prior and a Poissonian likelihood (see Kitaura \& En{\ss}lin 2008, Kitaura et al 2010). We assume a Gaussian prior $P(\delta(\{\mbi q\})|\mat C)$ describing the initial field $\delta(\{\mbi q\})$ which only depends on the correlation function (variance) $\mat C$ or power-spectrum and thus on a set of cosmological parameters. The likelihood  $P(\{\mbi q\}|\delta(\{\mbi q\})$ is modeled by a Poissonian PDF describing the discrete nature of the test particles at their Lagrangian positions. In order to sample from such a posterior PDF the efficient Hamiltonian Sampling technique is applied \citep[see][]{jasche_hamil,kitaura_lyman}. 
This step yields Gaussian fields on a mesh with $N_{\rm c}$ cells. One should note that similar approaches exist since long in the literature usually known under the term {\it constrained realisations}  \citep[see][]{bertschinger,hoffman,rien}.

\item {\bf Step 2 (Eq.~\ref{eq:structure}):   Structure matching step: {\color{black} Likelihood comparison:} Sampling the set of matter tracers at the initial conditions given an initial Gaussian field and a galaxy distribution.  }

The second step requires a structure formation model linking the initial Gaussian field to the observed structures given by the set of galaxies $\{\mbi x_{\rm G}\}$ to perform a {\it constrained simulation}. We need to sample the high dimensional statistical space spanned by the ensemble of nonlinearly evolved density fields compatible with the data.
To find computationally feasible solutions we have to assume at this stage a simplified structure formation model which remains accurate on large-scales. Motivated by the recent findings in the accuracy of tracing the cosmic structures back in time and estimating the peculiar velocity fields with second order Lagrangian perturbation theory (2LPT) we have chosen here to use this approximation \citep[see][]{kit_lin,kit_vel}. 
In this framework we can compute the displacement field $\mbi\Psi(\mbi q)$ from the Gaussian field  $\delta(\{\mbi q\})$ in an efficient way with FFTs. A random homogeneous Poisson distribution of  $N_p$ particles is moved to the redshift in which the observations are present according to the previously computed displacement field.  For technical details we refer to \citet[][]{1994MNRAS.267..811B,1995A&A...296..575B,2002PhR...367....1B}.
We then assign a number of closest particles $N^{\rm G}_p$ to each galaxy, where each particle must be closer than a certain distance $d_{\rm max}$.
Here we have done an efficient k-d tree implementation for different species based on  a homogeneous partition.  The id's of particles and galaxies are stored on a grid permitting us to look up fast which particles are in the surrounding of a particular galaxy.  
In this way each galaxy $\mbi x^i_{\rm G}$ is assigned a set of particles $\{\mbi x\}^i$ answering the question: where does the matter come from which formed the structures we observe today.
{\color{black}  The galaxy bias that is put in this method is given by the 2LPT halo bias constructed around galaxies \citep[for a review on the halo-model see][]{cooray-2002-372}. A similar procedure was suggested by \citet[][]{scocci}. This bias can be improved with a stochastic halo-model on top of the 2LPT realisation as performed in the same work. More sofisticated approaches could be done based on halo occupation distribution models \citep[see the recent work by][and references therein]{2012arXiv1203.6609M}. This point needs to be further investigated. One should notice that the first step searches for Gaussian fields compatible with a given power-spectrum and in this way already partly corrects for the bias in a $k$-dependent way.}
As we know the initial positions of the particles associated to each galaxy we obtain a distribution of Lagrangian positions $\{\mbi q\}$ which trace the initial Gaussian field. This closes one iteration permitting us to go to the first step again.
In the first iteration the Gaussian field is assumed to be zero and thus the initial displacement field vanishes.

\end{enumerate}

\begin{figure*}
\begin{tabular}{cc}
\hspace{0.cm}
\includegraphics[width=8.cm]{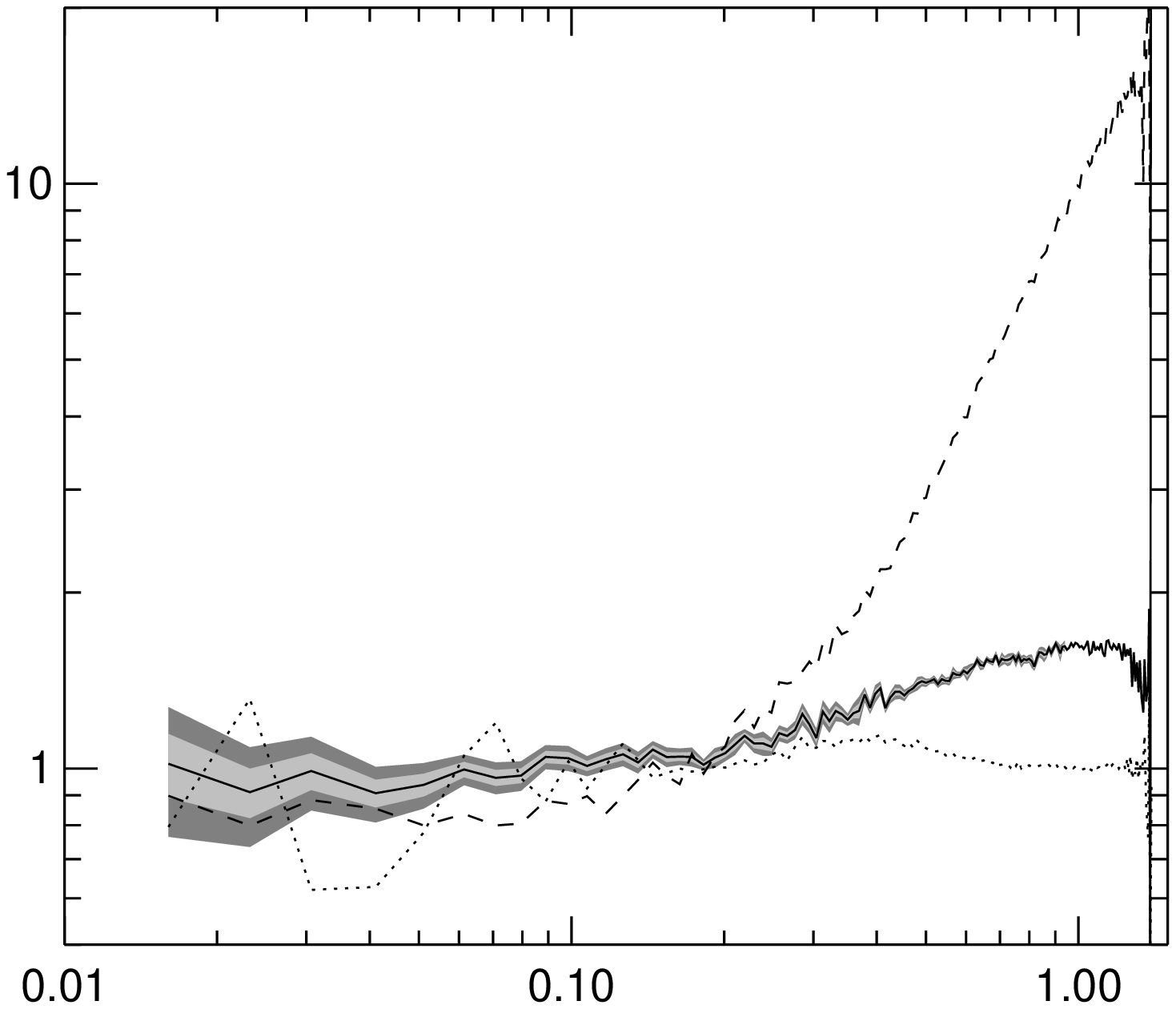}
\put(-235,100){\rotatebox[]{90}{$P(k)/P^{\rm L}(k)$}}
\put(-120,-5){$k$ [$h^{-1}$ Mpc]}
\hspace{-.5cm}
\includegraphics[width=8.cm]{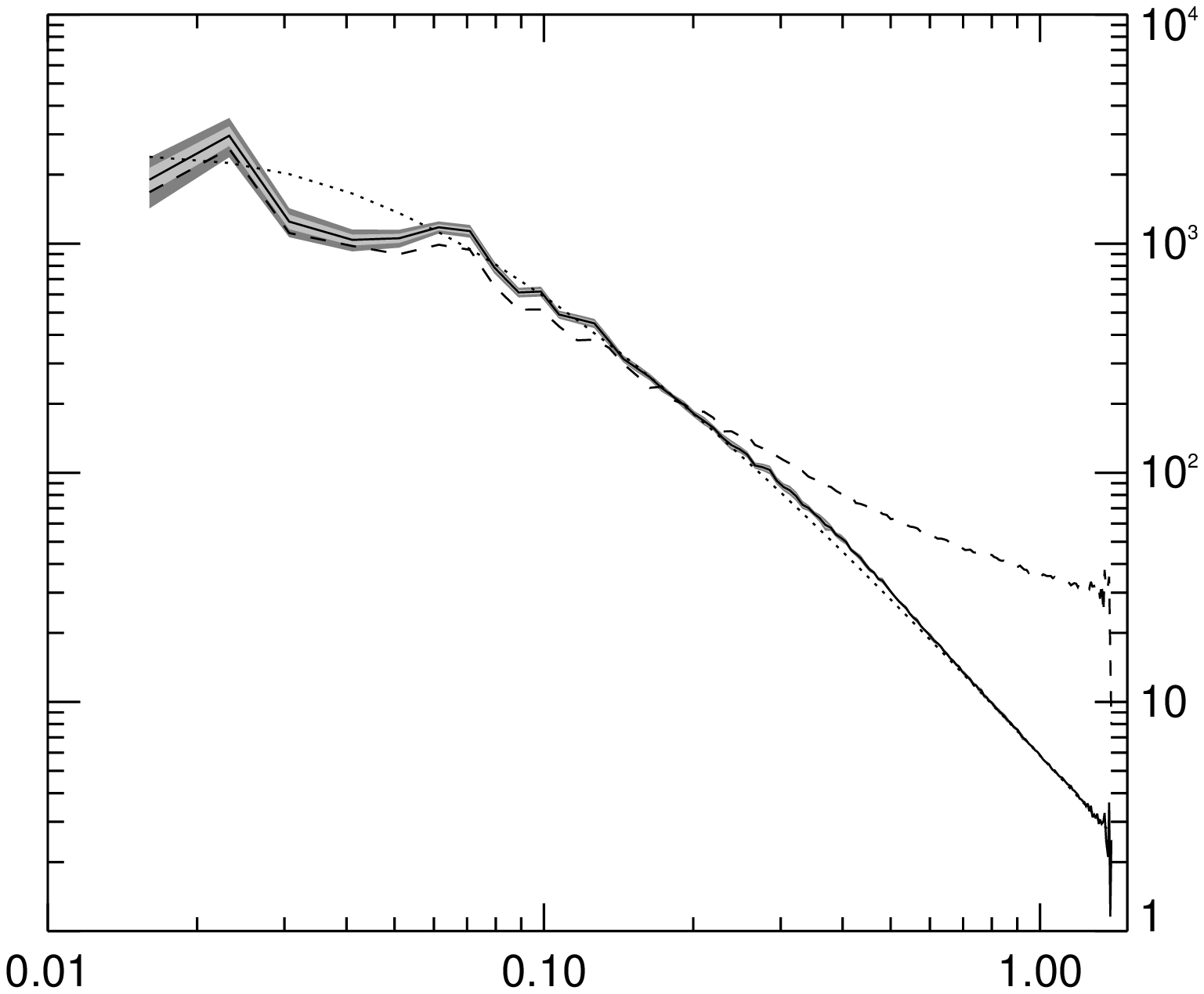}
\put(0,100){\rotatebox[]{-90}{$P(k)$}}
\put(-130,-5){$k$ [$h^{-1}$ Mpc]}
\end{tabular}
\caption{{\bf Left panel:} Ratio between the power-spectrum of the sample of 500 reconstructed initial Gaussian fields after 1000 iterations $P(k)$ and the power-spectrum from the first available snapshot of the Millennium Run at redshift $z=$127 gridded with nearest grid point on a mesh with $128^3$ cells ($P^{\rm L}(k)$) (black curve). The excess of power at high $k$ is due to aliasing. Dashed curve: same as previous case, but with power-spectrum $P(k)$ corresponding to the galaxy overdensity. Dotted curve: Same as continuous black curve, but with $P^{\rm L}(k)$ being given by the theoretical $\Lambda$-CDM model. {\bf Right panel:} power-spectra of the sample of 500 reconstructed initial Gaussian fields after 1000 iterations (black curve), the galaxy sample (dashed curve), and the theoretical $\Lambda$-CDM model (dotted curve). Shaded regions indicate 1 and 2 sigma contours.} 
\label{fig:pow}
\end{figure*}

\begin{figure*}
\vspace{-0.3cm}
\includegraphics[width=12.cm]{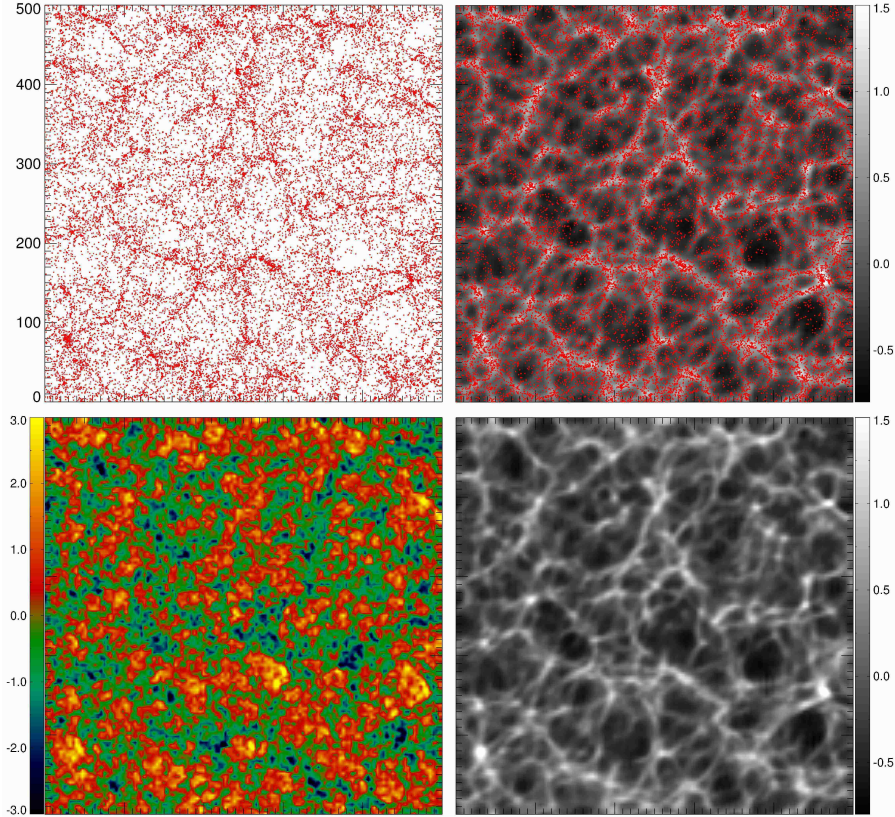}
\vspace{-0.3cm}
\caption{Reconstruction of the initial conditions and of the cosmic web. 
{\bf Upper left panel:} slice about 4 $h^{-1}$ Mpc thick averaged over 9 neighbouring slices through a galaxy catalogue (about 530 000 mock galaxies in a volume of 500 $h^{-1}$ Mpc side  \citep[][]{DeLucia-Blaizot-07}). Each galaxy is represented by a red circle. {\bf Upper right panel:} same slice through a sample after 1000 iterations showing {\color{black} the logarithm of the} reconstructed nonlinear matter density field using $384^3$ particles gridded on a $128^3$ mesh  with triangular shape cloud. The mock galaxies corresponding to the same slice are over-plotted indicating the accuracy of the reconstruction method.  {\bf Lower left panel:} 
same slice through the reconstructed initial conditions corresponding to the same sample which are Gaussian distributed. 
{\bf Lower right panel:} 
same as upper right panel without the mock galaxies. 
} 
\label{fig:rec}
\end{figure*}

\section{Validation of the method}

In this section the numerical experiments are presented which have been performed to validate the method.
As an input for our studies we have taken a galaxy catalogue  which uses a semi-analytic halo-model scheme \citep[][]{DeLucia-Blaizot-07}  based on the Millennium Run simulation \citep[][]{Springel-05} with a box of 500 $h^{-1}$ Mpc side. In particular we have considered a uniform subsample of about 530 000 galaxies including all galaxy types  conforming a nonlinear biased tracer of the underlying matter distribution.
This set-up permits us to test whether our reconstructions of the initial conditions and the cosmic web resemble the actual ones of the simulation.

We have made a great effort  optimising our computer code \textsc{Kigen} to cope with the hard  computational task presented in this work \citep[for the first version see][]{kit_lin}. We employ the Hamiltonian sampling technique in the Gaussianisation step based on the \textsc{Argo}-code \citep[first and last works][respectively]{kitaura,kitaura_lyman}  and the efficient 2LPT structure formation model with parallel  FFTs.

To compute the Gaussianisation step (Eq.~\ref{eq:shot}) and the displacement fields (Eq.~\ref{eq:structure})  with FFTs we have used a mesh of $N_{\rm c}$=$128^3$  cells. Structure formation was simulated using 2LPT with  $N_p$=$384^3$ particles. Only particles closer than $d_{\rm max}=2\,dL$ where considered to be {\it friends} of a galaxy with $dL=L/N_{\rm c}^{1/3}$ being the cell side, allowing for up to $N^{\rm G}_p=50$ particles associated to each galaxy. {\color{black} We have tested \textsc{Kigen}  with $N^{\rm G}_p=$1, 5, 10, 20, 30, 50, finding a stable behaviour already with $N^{\rm G}_p$=20. For safety we chose $N^{\rm G}_p$=50 leaving a thorough analysis for furture work.} We gave equal weights to those particles, but could improve the method considering different weights for instance according to the luminosity of the galaxy. We have also performed tests with fewer particles finding a worse resemblance with the actual density fields from the simulation. Nevertheless, a careful study is still to be done to quantify the required number and weights of particles for a given resolution and different galaxy type. In addition, we discarded cells with less or equal 3 particles as such weak constraints correspond to very low density regions which are not correctly captured by the 2LPT approximation. Going to third order could mitigate this problem \citep[see discussion in][]{scocci}.

We have performed 1500 iterations and found convergence in the matter statistics and the power-spectra of the reconstructed initial conditions after about 600 iterations. For safety we consider, however, only the last 500 iterations.


Let us start demonstrating that the first step in our reconstruction method leads to Gaussian distributed fields.  Fig.~\ref{fig:stats} shows the matter statistics of 500 samples after convergence has been achieved. 
The continuous black curve shows the mean PDF of the overdensity for the 500 samples of the reconstructed initial Gaussian field after 1000 iterations. The low skewness, kurtosis and mean values indicate that it is indeed Gaussian distributed, whereas the gravitationally evolved field using 2LPT unveils a highly skewed PDF (dashed curve).

Fig.~\ref{fig:pow} shows the power-spectrum of the reconstructed initial conditions in comparison to the one corresponding to the galaxy overdensity field and {\color{black} the ensemble averaged power-spectrum of primordial fluctuations in a  $\Lambda$-CDM Universe} with the same cosmological parameters as in the Millennium Run \citep[see][]{Springel-05}. The good agreement between the reconstruction and linear  theory demonstrates that shot noise and gravitational nonlinear evolution have been removed. On large scales ($k<0.2$ $h^{-1}$ Mpc) we find that the fields reveal a  great resemblance with the actual linear power-spectrum from the first snapshot of the simulation (see left panel of Fig.~\ref{fig:pow}). 

A visualisation of the results is given in Fig.~\ref{fig:rec}. Here we show how the reconstruction of the initial field leads to constrained nonlinear density field estimates which nicely follow the structures traced by the galaxies. This is emphasised by the upper right panel in which the input galaxies are overplotted on top of the cosmic web reconstruction. It is also remarkable how difficult it is to recognize the cosmic web in the initial Gaussian field (lower left panel in the  of Fig.~\ref{fig:rec}), and nevertheless how accurately this field leads to the observed structures.

\begin{figure}
\begin{tabular}{cc}
\hspace{-1.cm}
\includegraphics[width=6.cm]{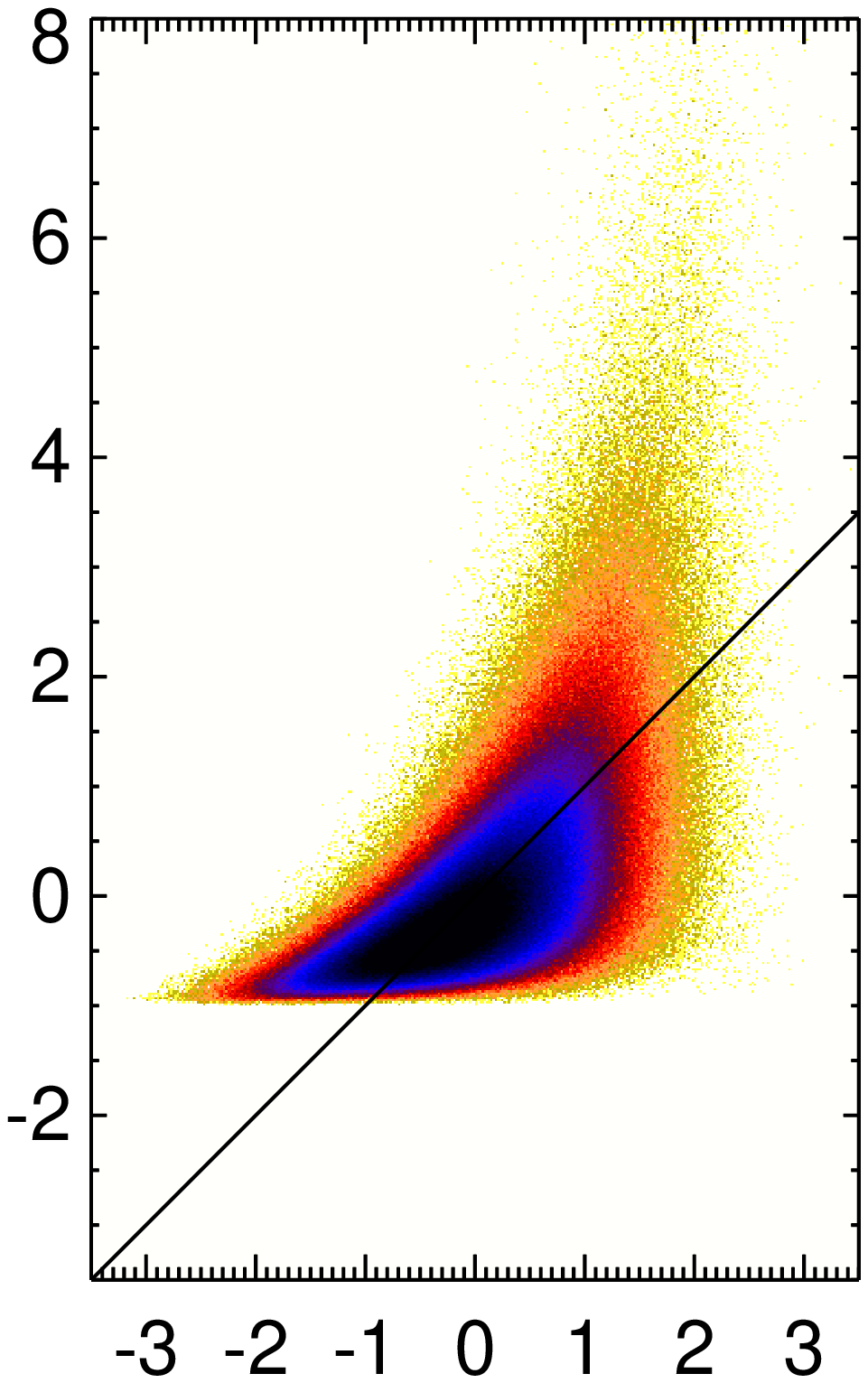}
\put(-145,90){\rotatebox[]{90}{$\delta^{\rm G}(\mbi x)$}}
\put(-80,-5){$\delta(\mbi q)$}
\hspace{-2.7cm}
\includegraphics[width=6.cm]{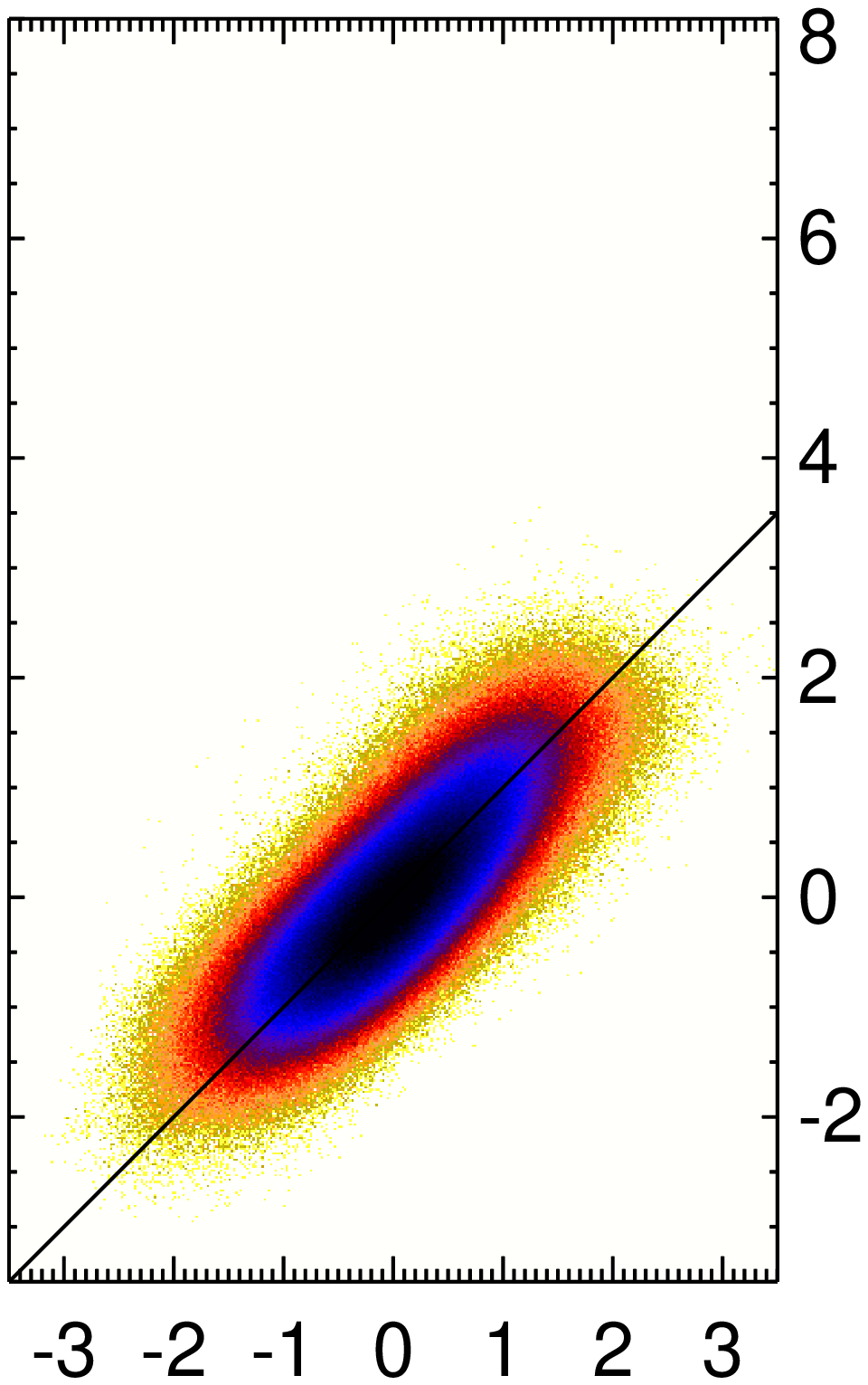}
\put(-15,90){\rotatebox[]{-90}{$\delta^{\rm rec}(\mbi q)$}}
\put(-80,-5){$\delta(\mbi q)$}
\end{tabular}
\caption{Cell-to-cell correlation after Gaussian smoothing with radius 5 $h^{-1}$ Mpc between: the initial conditions $\delta(\mbi q)$ given by the first available snapshot of the Millennium Run at redshift $z=$127 
({\bf left  panel:}) and the galaxy field $\delta^{\rm G}(\mbi x)$, 
({\bf right panel:}) and the reconstructed initial Gaussian field after 1000 iterations $\delta^{\rm rec}(\mbi q)$.
The dark colour-code indicates a high number and the light colour-code a low number of cells.} 
\label{fig:corr}
\end{figure}

To quantify this resemblance we have computed the cell-to-cell correlation between the initial conditions $\delta(\mbi q)$ given by the first available snapshot of the Millennium Run at redshift $z=$127  and  the galaxy field $\delta^{\rm G}(\mbi x)$ in comparison to our reconstructed initial field $\delta^{\rm rec}(\mbi q)$ after 1000 iterations (see upper panels in Fig.~\ref{fig:corr}, for similar studies see Kitaura and Angulo 2011).
We find that the complex nonlinear and nonlocal relation between the initial and final fields including galaxy biasing is straightened to a  closely unbiased relation with our reconstruction scheme. {\color{black} }

Finally we compute the propagator from the first available snapshot of the Millennium Run at redshift $z=$127 and the sample of reconstructed initial conditions after 1000 iterations and the galaxy overdensity, finding that the correlation is more than doubled at $k=$0.3 $h^{-1}$ Mpc.
We can also see however, that a number of galaxies in low density regions, mostly {\it living} in small filamentary structures, do not match any structure of the recovered cosmic web.
The reason is that those isolated galaxies which do not form groups do have a low number of particles associated to them which leaves them nearly unconstrained. We should also remind that the structure formation model we are using is simplified and especially fails in low density regions \citep[see][]{scocci}.

\section{Conclusions and Discussion}

In this work I have presented a new approach to recover the initial conditions  and the cosmic web structure underlying a galaxy distribution.
The method is based on sampling Gaussian fields which are constrained on  a galaxy distribution and a structure formation model. 

This is achieved by splitting the inversion problem into two Gibbs-sampling steps: the first being a Gaussianisation step transforming a distribution of point sources at Lagrangian positions --which are not a priori given-- into a linear alias-free Gaussian field. The second step being a matching procedure in which the set of matter tracers at the initial conditions is constrained on the galaxy distribution and the structure formation model we assume. For computational reasons we use second order Lagrangian Perturbation Theory. 
We demonstrate taking a  semi-analytic halo-model based galaxy mock catalog that the recovered initial conditions are closely unbiased with respect to the actual ones from the corresponding $N$-body simulation down to scales of a few $h^{-1}$ Mpc.
The cross-correlation between them shows a substantial gain of information.
In addition the initial conditions are extremely well Gaussian distributed and the power-spectra closely follow the shape of the linear power-spectrum down to scales of $k\sim1$ $h$ Mpc$^{-1}$.
The approach presented here has several advantages with respect to previous methods. It provides a set of joint estimates of the initial conditions, the cosmic  web and the peculiar velocity field given a galaxy distribution.

 {\color{black} We note, that during the submission process of this letter an independent work has appeared on the arxiv trying to solve the same problem presented here using the second order Lagrangian perturbation framework within Eq.~\ref{eq:step0} \citep[][]{jasche_2lpt}. 
 Their method is substantially different from the one presented here, although in its essence both try to solve the boundary problem of finding the initial conditions in a probabilistic way using the Bayesian framework.   
 While the present approach uses an adaptive particle based scheme and hereby models the collapse of Lagrangian regions to compact objects following the idea suggested in \citet[see][]{scocci} and developed in \citet[][]{2012arXiv1203.6609M}, their grid-based approach relies on the Poissonian assumption to model the galaxy distribution. Furthermore their approach has been only tested with mock catalogues generated with such a 2LPT-Poisson model while we test ours with a semi-analytic halo-model based galaxy sample based on $N$-body simulations.
 The \textsc{Kigen}-code can be immediately substituted by a different structure formation model, while their's is more focused on 2LPT leaving the redshift distortions unresolved. We note that this can be modeled in a straightforward way within the particle based method presented here. One just needs to add a second displacement to include coherent flows having the following mapping relation $\mbi s=\mbi q+\mbi \Psi+(\mbi v\cdot\hat{\mbi r})\hat{\mbi r}/(Ha)$, where $\mbi v$ is the full three dimensional velocity field \citep[which can be accurately computed within 2LPT, see][]{kit_vel}, $\hat{\mbi r}$ is the unit sight line vector, $H$ the Hubble constant and $a$ the scale factor. Fingers-of-god could be modeled with a dispersion term \citep[see discussion in][]{kitaura}. }
 The \textsc{Kigen}-code is flexible and could be extended to be more precise by including a more detailed structure formation description like the one provided by full $N$-body simulations. This will imply, however, a high computational cost.
Although the results seem to indicate that this approach could be useful for power-spectrum reconstructions as well, one should be cautious since the power-spectrum is included as a prior in the Gaussianisation step. This assumption could be relaxed including an additional Gibbs-sampling step for power-spectrum estimation as proposed by \citet[][]{kitaura,jasche_gibbs,kitaura_lyman}. Note that the assumed power-spectrum did not include any kind of wiggles. The features in the power-spectra of the recovered fields show however a great resemblance with the actual one indicating that this method could be useful for BAO reconstruction.
 Redshift space distortions caused by the proper motions of galaxies can be treated in the same framework as we show in a subsequent work \citep[see discussion above and][for a similar approach]{kitaura_lyman}.  
The scatter between the recovered initial conditions and the actual ones could be reduced by improving the resolution and the constraints, for instance giving different weights to the galaxies according to their luminosity, or using a compiled halo catalogue as an input.
 This framework could also permit one to include a  light-cone and consider hereby evolution effects by producing different snapshots at several redshifts. 
The forward approach is also especially robust as it is calibrated  with the input data in each iteration. 

In summary, the method presented here provides a very powerful approach to the inversion problem of density field reconstruction of the initial conditions and the present cosmic web.

\section*{Acknowledgments}

FK thanks Raul Angulo, Volker M\"uller, Arman Khalatyan, Stefan Gottl\"ober, Yehuda Hoffman, Timur Doumler, Simon D.~M.~White, Bjoern Malte Schaefer, Sebastian Nuza, Steffen Hess, and Ravi Sheth for encouraging discussions. This work has been presented at the {\it cosmic flow} conference in Australia the 21st of February 2012.

{\small
\bibliographystyle{mn2e}
\bibliography{lit}
}

\end{document}